\documentclass[a4paper,12pt]{article}
\usepackage{epsf}

\def\lsim{\mathrel{\rlap{
\lower4pt\hbox{\hskip-3pt$\sim$}}
    \raise1pt\hbox{$<$}}}     
\def\gsim{\mathrel{\rlap{
\lower4pt\hbox{\hskip-3pt$\sim$}}
    \raise1pt\hbox{$>$}}}     
\begin{document}
\title{Dynamical Interpretation  \\
   of Chemical Freeze-Out in Heavy Ion Collisions }
\author{V.D.~Toneev$^{1,2}$, J. Cleymans$^3$,
E.G.~Nikonov$^{1,2} $, K.~Redlich$^{1,4}$,\\
and \\
A.A.~Shanenko $^{2}$\\
\vspace*{2 mm}
\small\em
$^1$Gesellschaft f\"{u}r  Schwerionenforshung, D-64220  Darmstadt, Germany \\
\small\em
$^2$ Bogoliubov Laboratory of Theoretical Physics,\\
\small\em
JINR, 141980 Dubna, Moscow Region,  Russia\\
\small\em
$^3$ Department of Physics, University of Cape Town, Rondebosch 7701,
South Africa\\
\small\em
$^4$ Institute of Theoretical Physics, University of Wroclaw, PL-64291
Wroclaw, Poland.
}
\date{}
\maketitle

\begin{abstract}
It is demonstrated that there exists  a direct correlation
between chemical freeze-out point and the softest point of the
equation of state where the pressure divided by the energy
density, $p(\varepsilon)/\varepsilon$, has a minimum. A dynamical
model is given as an example where  the passage of the softest
point coincides with the condition for chemical freeze-out,
namely an average energy per hadron $\approx$ 1 GeV. The
sensitivity of the result to the equation of state used is
discussed.
\end{abstract}


%
%

\newpage

Over the last few years the question of chemical equilibrium in
heavy ion collisions has attracted much attention
\cite{BM,SSG97,sollfrank,BWE}. Assuming thermal and chemical
equilibrium within a statistical model, it has now been shown that
it is indeed possible to describe the hadronic abundances
produced  at beam energies ranging
 from 1 to 200 AGeV. The observation was made that the
 parameters of the  chemical freeze-out curve
obtained at CERN/SPS, BNL/AGS and GSI/SIS all lie on a unique
freeze-out curve in the  temperature $T$ -- chemical potential
$\mu_B$ plane~\cite{BMS96}. Recently, a surprisingly simple
interpretation of this curve has been proposed: the hadronic
composition at the final state is determined solely by the
average energy per hadron,
$\left<E_{had}\right>/\left<N_{had}\right>$, of approximately 1
GeV  in the rest frame of the system under consideration
\cite{CR98,CR99}. Using this finding
$\left<E_{had}\right>/\left<N_{had}\right> \approx 1$ GeV as a
heuristic definition for the {\em chemical freeze-out point}, in
this letter we study physical conditions which are realized
around this point in evolution of the system. We  show that  the
chemical freeze-out point  is intimately related to {\em the
softest point} of equation of state defined by  the minimum of
the pressure-to-energy density ratio, $\displaystyle
p(\varepsilon)/\varepsilon$, as a function of
$\varepsilon$~\cite{HS94}.

 Our considerations are essentially based on the
 recently proposed
Mixed Phase (MP) model~\cite{NST98,TNS98}  which is consistent
with the available QCD lattice data \cite{karsch}. The underlying
assumption of the MP model is that  unbound quarks and gluons
{\it may coexist} with hadrons forming an {\it homogeneous}
quark/gluon--hadron phase \cite{NST98,TNS98}. Since the mean
distance between hadrons and quarks/gluons in this mixed phase
may be of the same order as that between hadrons, their
interaction with unbound quarks/gluons plays an important role
defining
 the order of the phase transition. Recently the importance of
quark/gluon--hadron interactions was also discussed in the context
of dilepton production~\cite{TLM99}.

Within the MP model \cite{NST98,TNS98} the
effective Hamiltonian is written in the quasiparticle
approximation with the density-dependent mean--field interaction.
 Under quite general requirements of  confinement
for color charges,
the mean--field potential of  quarks and gluons
is approximated by the following form:
\begin{equation}
U_q(\rho)=U_g(\rho)={A\over\rho^{\gamma}}
\label{eq6}\end{equation}
with  {\it the total density of quarks and gluons}
$$
\rho=\rho_q + \rho_g +\sum\limits_{j}\;n_j\rho_{j}
$$
where $\rho_q$ and  $\rho_g$ are the densities of unbound quarks
and gluons outside hadrons, while $\rho_{j}$ is the density and
$n_j$ is the number of valence quarks inside the hadron of type
$j$. The presence of the total density $\rho$ in (\ref{eq6})
describes  the interaction between all components of the mixed
phase. The  approximation (\ref{eq6})  recovers two important
limiting cases of the QCD interaction. Namely, if $\rho
\rightarrow 0$ the interaction potential goes to infinity, {\it
i.e.} an infinite energy is required to create an isolated quark
or gluon  which ensures the confinement of color objects. In the
other extreme  of high energy density corresponding to $\rho
\rightarrow \infty$, we obtain the asymptotic  freedom regime.

The use of the density-dependent  potential (\ref{eq6}) for quarks
 and the hadronic potential described by a
modified non-linear mean--field model~\cite{Zim}  requires certain
constraints  to satisfy thermodynamic
consistency~\cite{NST98,TNS98}. For the
chosen form of the Hamiltonian these conditions
show that $U_g(\rho)$ and
$U_q(\rho)$ should be independent of the temperature. From
these requirements one also obtains an  expression for
 the quark--hadron potential~\cite{NST98}.  In the case when
  the quark-gluon component is  neglected, the MP model is reduced
  to the interacting hadron gas model with a non-linear mean--field
  interaction~\cite{Zim}.

A detailed study of the pure gluonic medium
with  $SU(3)$ color symmetry and a first
order phase transition allows to fix the values of
 $\gamma =0.62$ and $\displaystyle
A^{1/(3\gamma+1)} = 250\, MeV$. These values are then generalized
to the  $SU(3)$ system with dynamical quarks. For two light
flavors at zero baryon density, $n_B=0$, the MP model is
consistent with the results from lattice QCD \cite{karsch}  with
the deconfinement
 temperature $T_{dec}=153\;MeV$ and the crossover type of
 phase transition. The model is then proposed to be  extended to
baryon-rich  systems in a parameter--free way \cite{NST98}.

A particular consequence of the MP model is that for $n_B=0$
 the softest  point~\cite{HS94} of the equation of state
  is located at comparatively  low values of the
energy density: $\varepsilon_{SP} \approx 0.45 \ GeV/fm^3 $ which
is in a good agreement with recent lattice
estimates~\cite{karsch}. This value of $\varepsilon_{SP} $ is
close to the energy density inside a nucleon. Thus, reaching this
value signals  that we are dealing with  a single 'big' hadron
consisting of deconfined matter. For baryonic matter the softest
point is gradually washed out at $n_B \gsim 0.4 \ n_0$. As shown
in~\cite{NST98,TNS98}, this  thermodynamic behavior differs
 drastically  from both the interacting
hadron gas model which has  no  softest point  and the two--phase
approach, based on the bag model, having  by construction a first
order phase transition and the softest point at $\varepsilon_{SP}
> 1 \ GeV/fm^3 $   independent of $n_B$~\cite{HS94}. These
differences should manifest themselves in the expansion dynamics.

In Fig.1 we show   the evolution trajectories for central $Au+Au$
collisions in the $T-\mu_B$ plane  together with the freeze-out
parameters obtained from hadronic abundances. Using  the
quark-gluon string model~\cite{QGSM}, the initial state was
estimated as a state of hot and dense nuclear matter inside a
cylinder in the center-of-mass frame with radius $R=4 \ fm$ and
length $L=2R/\gamma_{c.m.}$.  The calculated temporal behavior of
energy and baryon densities is close to that in RQMD or UrQMD
transport models and the procedure of defining the initial state
of a fireball is described in~\cite{NST98,TNS98}. The subsequent
isentropic expansion of this fireball is treated within the
Frankfurt expansion model~\cite{SSW86} based on a scaled
hydrodynamical prescription assuming $t \sim V$. As seen in
Fig.1, the {\em turning points} of these trajectories,  {\it
i.e.} the points where $\partial T /\partial \mu_B$ changes the
sign, are in a good agreement
 with the extracted  chemical freeze-out parameters and
 clearly correlate with the smooth curve for the ideal gas model
with $\left<E_{had}\right>/\left<N_{had}\right>=1$ GeV
 \cite{CR98} defined above as a condition of  chemical freeze-out.
Existence of the turning point is related to two limiting
equilibrium regimes for temperature behavior of the chemical
potential: ultrarelativistic regime dominated by light particles
(quarks, pions) when $\mu_B \sim T^2$ and non-relativistic one
defined mainly by nucleons and deltas
with $\mu_B \sim T^{-1}$ \\
- - - - - - - - - - - - - - - - -  - - - - - - - - - - - - - - - - - - - \\
{\it Footnote:} Other approach have been recently  applied
in~\cite{Bra99} where conditions for thermodynamic equilibrium in
heavy ion collisions were studied within the transport UrQMD
model and the calculation results at a fixed time moment were
approximated by the ideal hadron gas equation of state to extract
the time dependence of $T$  and $\mu_B$. It is of interest that
our results for dynamical trajectories below the turning point
practically coincide with that from~\cite{Bra99} for {\em
equilibrium} part of $T-\mu_B$ trajectory. However, at higher
temperature, where according to~\cite{Bra99} the system is in a
{\em non-equilibrium} state,
their results differ from ours and exhibit no turning point.\\
- - - - - - - - - - - - - - - - -  - - - - - - - - - - - - - - - - - - - \\
. Uncertainty in the initial $\varepsilon$ and $n_B$ in 10-20$\%$
shifts slightly the dynamical trajectory but the turning point
nevertheless stays on the
$\left<E_{had}\right>/\left<N_{had}\right>=1$ GeV curve. Notice
that there is a finite region in $T$ where $\partial
\mu_B/\partial T \approx 0$,  {\it i.e.} the chemical potential
in the expanding system is kept constant around this turning
point.

The observed correlation is further elucidated in Fig.2. The quantity
$p/\varepsilon$ is closely related
to the square of the velocity of sound  and
characterizes the expansion speed \\
- - - - - - - - - - - - - - - - -  - - - - - - - - - - - - - - - - - - - \\
{\it Footnote:} In  simplified hydrodynamic models was shown
that, for example, the transverse expansion of a cylindrical
source  is governed by the pressure-to-enthalpy ratio,
$p/(p+\varepsilon)$~\cite{CGS86,rischke}, {\it i.e.} by
$p/\varepsilon$ rather than  by the sound velocity. \\
- - - - - - - - - - - - - - - - -  - - - - - - - - - - - - - - - - - - - \\
, so the system lives for an appreciable amount of time around the
softest point which  facilitates to reach the chemical
equilibrium. This statement is evidenced directly in the
right-hand side of Fig.2 where the time evolution of
$p/\varepsilon$ is shown: The system spends about 1/3 of the
total  expansion time in a state near the softest point. During
this time the baryon chemical potential $\mu_B$ remains
practically constant.

In Fig.2 (left-hand side) the position of the softest point
correlates with the average energy per hadron being about  1 GeV
for all beam energies except for 2 AGeV case. One should note
that the quantity $\varepsilon /\rho_{had}$, where $\varepsilon$
is the total energy density, coincides with
$\left<E_{had}\right>/\left<N_{had}\right>$ considered
in~\cite{CR98} only in the case when there are no unbound
quarks/gluons in the system. In the MP model, all components are
interacting with each other and  therefore the quantity
$\left<E_{had}\right>$ is not well defined. However the admixture
of unbound quarks at  the softest point $\varepsilon_{SP}$
amounts to about  $13\%$ and $8\%$ at  beam energies
$E_{lab}=150$ and $10 \ AGeV$, correspondingly. Thus
thermodynamical quantities and in particular the {\em ratios} of
hadron abundance in the MP and resonance gas model are very close
to each other at the freeze-out point.

  The MP equation of state plays a decisive role for the
regularity considered
here, describing both the order of the phase transition
and the deconfinement temperature.  The
two-phase (bag) model exhibits a first order phase transition with
 $T_{dec}=160 \ MeV$ and has a {\it spatially
separated}  Gibbs mixed phase but the corresponding trajectories in the
$T-\mu_B$ plane are quite different from those in the MP
model as shown in~\cite{TNS98}.
The exit point from the Gibbs mixed phase at
$E_{lab}=150$  AGeV is  close to the corresponding
freeze-out point in Fig.1. However it  does not
lead to the observed correlation with the softest point position
in the whole energy range considered.
The interacting hadron gas model has no softest point effect as was
 demonstrated in~\cite{NST98,TNS98}  but nevertheless it exhibits the
turning point in $T-\mu_B$ plane.  These facts are seen also from Fig.1 and
Fig.2  where for $E_{lab}=2 \ AGeV$  quark-hadron interactions are
negligible ($\approx 1\%$). Indeed, at this energy instead
of a minimum  one has a monotonic fall-off specific for hadronic models
with only a small irregularity in $p/\varepsilon$  near the
point $\varepsilon /\rho_{had} \sim 1 \ GeV$ \\
- - - - - - - - - - - - - - - - -  - - - - - - - - - - - - - - - - - - - \\
{\it Footnote:} Note that at the SIS
energies the chemical freeze-out point practically coincides with the thermal
freeze-out~\cite{oeschler}.\\
- - - - - - - - - - - - - - - - -  - - - - - - - - - - - - - - - - - - - \\
.  At higher energies the dynamical trajectories
 are very different for hadronic and mixed phase models but the positions
of turning points in the $T-\mu_B$ plane turn out to be still
rather close to each other and
 both  lying on the  curve corresponding to $\varepsilon /\rho_{had}=
 1 \ GeV$. Therefore,  one could expect similar results for
the {\em ratios} of hadronic abundance in these two models.
However the  model predictions differ essentially as long as
evolution of the system is concerned.

It is noteworthy that similarly to the results presented in Fig.2,
the softest point of the equation of state correlates with an
average energy per quark, $\varepsilon/\rho \approx 350 \ MeV$
which is close to the constituent quark mass. So, at higher
values of $\varepsilon/\rho$ we are dealing with a
strongly-interacting mixture  of highly-excited hadrons and
unbound massive quarks/gluons forming (in accordance with
Landau's idea~\cite{landau}) an 'amorphous' fluid  suitable for
hydrodynamic treatment. Below the  softest point the interaction
deceases, the relative fraction of unbound quarks/gluons  is
getting negligible, higher hadronic resonances decay into baryons
and light mesons. At this stage the application of hydrodynamics
becomes questionable.

In summary, we have shown in this paper an example of an equation
of state where  there  exists a direct correlation between
chemical freeze-out point and the softest point of the equation
of state. This description correlates with the observation that
chemical freeze-out  occurs when the average  energy per hadron,
$\left<E_{had}\right>/\left<N_{had}\right>$, drops below the
value of 1 GeV as found in~\cite{CR98}. In the examples
considered such a clear correlation is observed for the  equation
of state of the mixed phase model, where the energy per hadron
seems to be a relevant observable defining chemical freeze-out
conditions.

\vspace*{5mm} We thank B.~Friman,  Yu.~Ivanov, E.~Kolomeitsev and
W.~N\"{o}renberg for useful discussions. E.G.N., K.R. and V.D.T.
gratefully acknowledge the hospitality at the Theory Group of
GSI, where some part of this work has been done. J.C. gratefully
acknowledges the hospitality of the Physics Department of the
University of Bielefeld. This work was  supported in part by DFG
under the program of scientific-technological collaboration
(project 436 RUS 113/558/0) and by Russian Found of Fundamental
Investigations (project 00-02-04012). K.R also acknowledges the
partial support of the State Committee for Scientific Research
(KBN).


\newpage
\begin{figure}[htb]
\begin{center}
\leavevmode
\epsfxsize=12.cm
\epsfbox{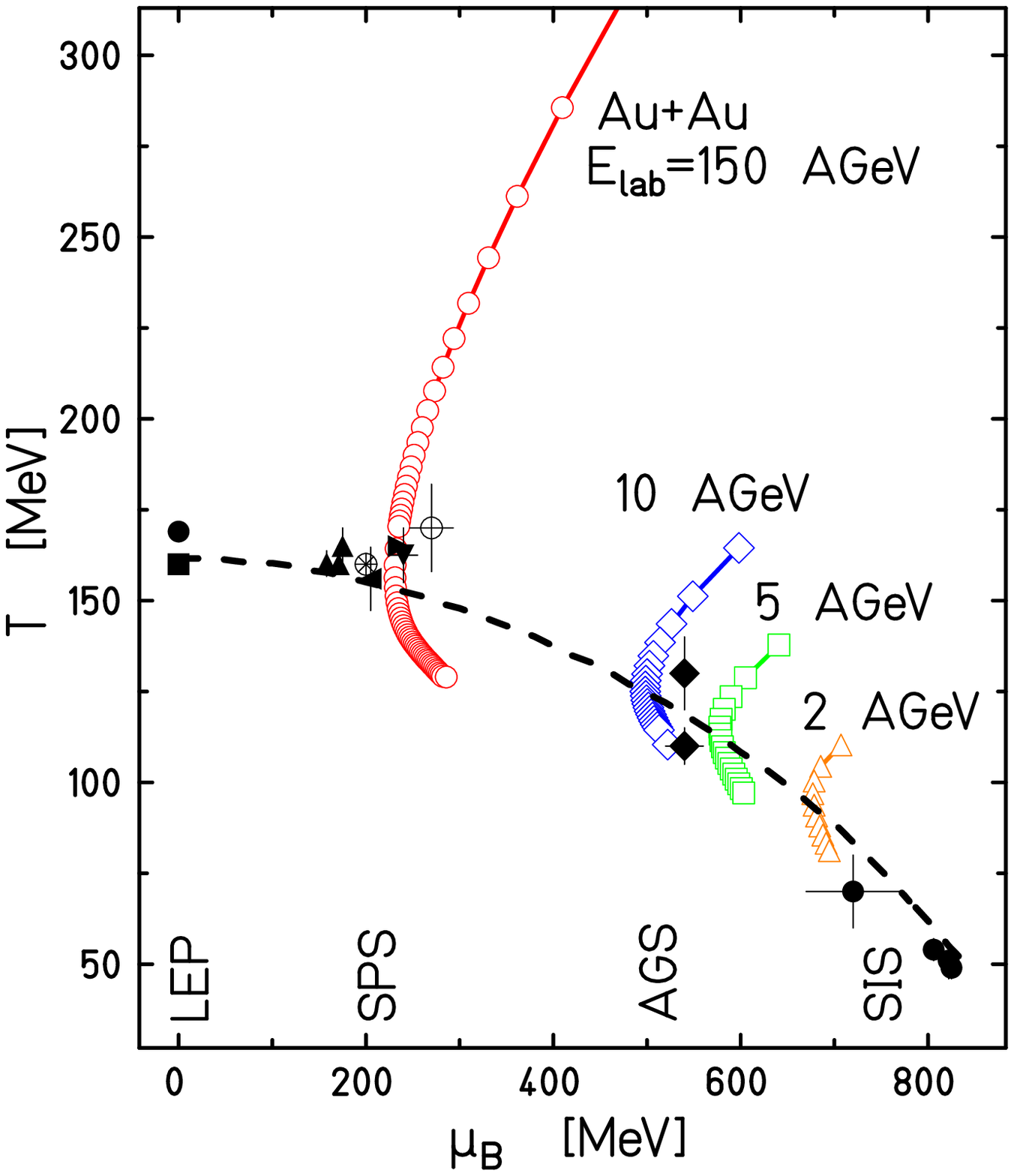}
\end{center}
\caption[C1]{ The compiled chemical freeze-out parameters
(from~\cite{CR98,CR99}) obtained from the observed hadronic
abundances and dynamical trajectories calculated for central
$Au+Au$ collisions at different beam energies $E_{lab}$  with the
mixed phase equation of state. The smooth dashed curve is
calculated in the  ideal hadron  gas model for
$\left<E_{had}\right>/\left<N_{had}\right> = 1 \ GeV$
~\cite{CR98}. }
  \label{fig1}
\end{figure}

\newpage
\begin{figure}[htb]
\begin{center}
\leavevmode
\epsfxsize=15.cm
\epsfbox{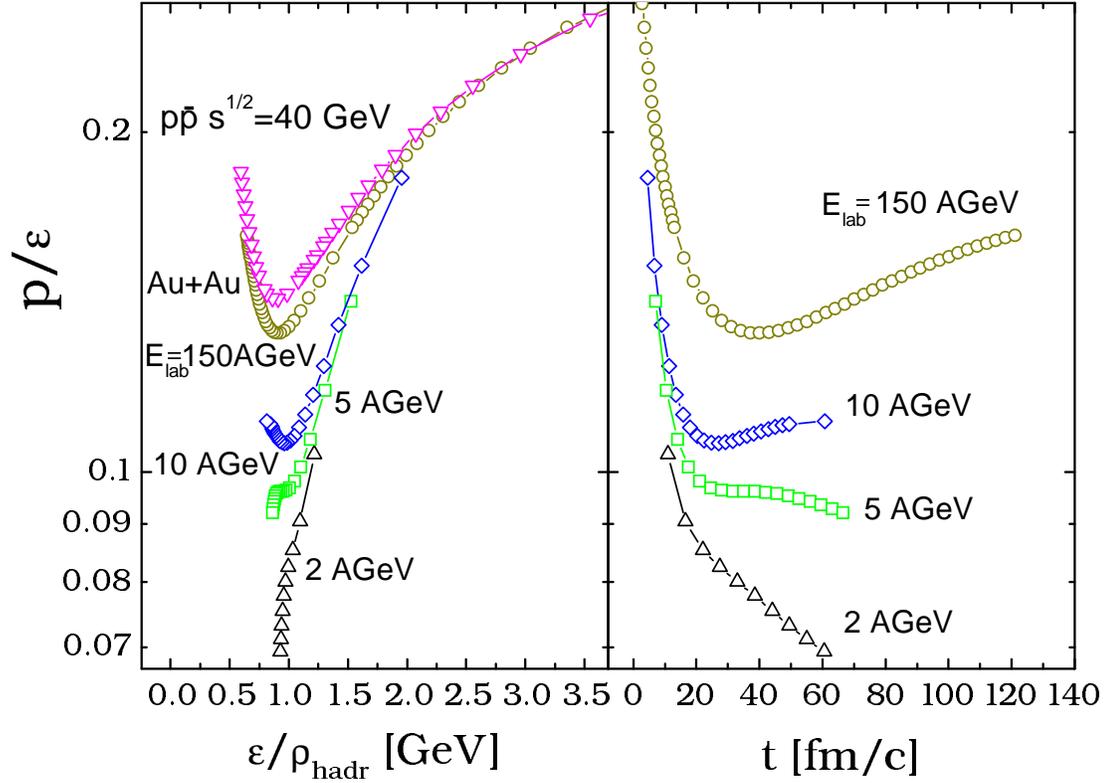}
\end{center}
\caption[C2]{ The ratio of pressure to energy density, $p/\varepsilon$, versus
the average energy per hadron, $\varepsilon/\rho_{had}$, (left-hand side)
and  its time evolution for different systems (right-hand side).  The
 curve for $p\bar p$ collisions at $\sqrt{S}=40 \ GeV$ is calculated for
 isentropic expansion of a sphere with $R=1 \ fm$. Other cases
are calculated for central $Au+Au$ collisions at the given beam
energy and under the same conditions as in Fig.1.
}
  \label{fig1}
\end{figure}

\end{document}